\documentclass[11pt]{amsart}
\usepackage[foot]{amsaddr}
\usepackage[T2A]{fontenc}
\usepackage[cp1251]{inputenc} 
\usepackage[russian,english]{babel}
\usepackage{amsfonts,amssymb,mathrsfs,amscd,comment,amsthm}

\usepackage{cite}
\usepackage{arydshln}
\usepackage{tcolorbox}
\usepackage{tikz}
\usepackage{rotating}
\usetikzlibrary{automata, positioning, arrows, calc,shapes}

\usepackage[final]{hyperref}

\hypersetup{
 colorlinks = true, 
 urlcolor  = blue, 
 linkcolor = blue, 
 citecolor = blue 
}

\DeclareSymbolFont{rsfscript}{OMS}{rsfs}{m}{n}
\DeclareSymbolFontAlphabet{\mathrsfs}{rsfscript}

\DeclareMathOperator{\dt}{.}

\DeclareMathOperator{\rt}{\mathrm{rt}}

\DeclareMathOperator{\excl}{\mathrm{excl}}
\DeclareMathOperator{\dupl}{\mathrm{dupl}}
\DeclareMathOperator{\dist}{\mathrm{dist}}

\newcommand{\sa}{synchronizing automata}
\newcommand{\san}{synchronizing automaton}

\newcommand{\sw}{reset word}
\newcommand{\sws}{reset words}
\newcommand{\ssw}{reset word of minimum length}
\newcommand{\rl}{reset threshold}
\newcommand{\scc}{strongly connected component}
\newcommand{\scn}{strongly connected}

\newcommand{\mA}{\mathrsfs{A}}

\newcommand{\mC}{\mathrsfs{C}}
\newcommand{\mD}{\mathrsfs{D}}
\newcommand{\mR}{\mathrsfs{R}}

\newtheorem{proposition}{Proposition}

\newtheorem{lemma}{Lemma}

\makeatletter
\newcommand{\l@abcd}[2]{\hbox to\textwidth{#1\hfill\textbf{#2}}}
\makeatother

\def\Ananichev{Ананичев}
\def\Cerny{\v{C}ern\'y}
\def\Rystsov{Рысцов}
\def\Ristsov{Рисцов}
\def\Pribavkina{Прибавкина}

\begin{document}

\title[Results on the \v{C}ern\'y Conjecture]{List of Results on the \v{C}ern\'y Conjecture and\\ Reset Thresholds for Synchronizing Automata}

\author[M. V. Volkov]{Mikhail V. Volkov}
\address{{\normalfont 620075 Ekaterinburg, Russia}}
\email{m.v.volkov@urfu.ru}

\date{}

\thanks{Supported by the Ministry of Science and Higher Education of the Russian Federation, project FEUZ-2023-2022}

\begin{abstract}
We survey results in the literature that establish the \v{C}ern\'y conjecture for various classes of finite automata. We also list classes for which the conjecture remains open, but a quadratic (in the number of states) upper bound on the minimum length of reset words is known. The results presented reflect the state of the art as of \today.
\end{abstract}

\maketitle

\tableofcontents

\section*{Introduction}
\label{sec:intro}

The author's survey~\cite{Volkov:2022} included an extensive section listing known partial results on the \v{C}ern\'y conjecture. While the list was reasonably complete at the time the survey appeared in print (September 2022), it has since become outdated as new results continue to emerge. Such is the inevitable fate of any `snapshot' of an active research area, once it is fixed in a journal publication. This has led the author to the idea of using the service provided by arXiv to maintain a living list of results on the \v{C}ern\'y conjecture and reset thresholds for synchronizing automata. New results will be appended to the list shortly after their appearance in the literature.

Section~\ref{sec:prelim} introduces the basic concepts and methods of the area, to the extent necessary for understanding and discussing the results included in our list. The list itself is presented in Section~\ref{sec:results}. The bibliography provides sources for each result in the list. As a rule, we omit preliminary publications (such as preprints and conference proceedings) that have been superseded by later journal articles, retaining only those of historical interest.

Of the results mentioned in the list, two have not yet appeared in the literature; their proofs are provided in the Appendix.

\section{Concepts and methods}
\label{sec:prelim}

\subsection{Basic definitions}
A \emph{complete deterministic finite automaton} (a DFA, for short) is a triple $\mA=\langle Q,\Sigma,\delta\rangle$, where $Q$ and $\Sigma$ are finite non-empty sets and $\delta\colon Q\times\Sigma\to Q$ is the \emph{transition function}. The elements of $Q$ and $\Sigma$ are referred to as \emph{states} and (\emph{input}) \emph{letters}, respectively, so $Q$ is the \emph{state set} and $\Sigma$ is the \emph{input alphabet} of $\mA$. Automata are understood to evolve in discrete time: at each moment, a DFA $\mathrsfs{A}=\langle Q,\Sigma,\delta\rangle$ is in a certain state $q\in Q$. During the next time unit, exactly one input letter $a\in\Sigma$ arrives, causing $\mA$ to transition to the state $\delta(q,a)$.

A DFA $\mathrsfs{A}=\langle Q,\Sigma,\delta\rangle$ can be represented by a labeled directed graph with the vertex set $Q$, the label alphabet $\Sigma$, and the set of edges
\[
\{q\xrightarrow{a}q'\mid q,q'\in Q,\ a\in\Sigma,\ \delta(q,a)=q'\}.
\]
Thus, the transition from state $q$ to state $q'$ under input $a$ is represented by an edge from $q$ to $q'$ labeled by $a$. The \emph{underlying graph} of $\mA$ is obtained by omitting the labels from this graph representation.
A DFA is called \emph{strongly connected} if its underlying graph is strongly connected.

A \emph{word} over an alphabet $\Sigma$ is any finite sequence of letters from $\Sigma$. The set of all words over $\Sigma$, including the \emph{empty word} $\varepsilon$, is denoted by $\Sigma^*$. If $w=a_1\cdots a_\ell$ with $a_1,\dots,a_\ell\in\Sigma$ is a non-empty word, the number $\ell$ is said to be the \emph{length} of $w$ and is denoted by $|w|$. The length of the empty word is defined to be 0. Words are multiplied by concatenation, and the empty word serves as the identity element for this multiplication. Thus, $\Sigma^*$ is a monoid.

The transition function $\delta$ of $\langle Q,\Sigma,\delta\rangle$ extends to a function $Q\times\Sigma^*\to Q$ (still denoted by $\delta$) in the following way. For each $q\in Q$, let $\delta(q,\varepsilon):=q$ and if $w=a_1\cdots a_\ell$ with $a_1,\dots,a_\ell\in\Sigma$ is a non-empty word, let
\[
\delta(q,w):=\delta(\dots\delta(\delta(q,a_1),a_2),\dots,a_\ell).
\]

For any DFA $\mathrsfs{A}=\langle Q,\Sigma,\delta\rangle$, every word $w\in\Sigma^*$ induces the transformation $q\mapsto\delta(q,w)$ on the set $Q$. The set of all transformations induced this way is closed under the composition of transformations and contains the identity transformation (induced by the empty word). Thus, it constitutes a monoid called the \emph{transition monoid} of $\mA$.

When a DFA is fixed, one often simplifies the notation by suppressing the transition function symbol. That is, the DFA is specified as a pair $\langle Q,\Sigma\rangle$, and one writes $q \dt w$ instead of $\delta(q, w)$.

\subsection{Synchronizing automata and the \v{C}ern\'{y} conjecture}
A DFA $\mA=\langle Q,\Sigma\rangle$ is called \emph{synchronizing} if there exist a word $w\in\Sigma^*$ and a state $s \in Q$ such that $q\dt w = s$ for all $q\in Q$. Such a word $w$ is said to \emph{reset $\mA$ to the state $s$} and is called a \emph{reset word} for $\mA$. The minimum length of reset words for $\mA$ is called its \emph{reset threshold} and denoted by $\rt(\mA)$.

Synchronizing automata are the subject of a rich theory with diverse connections and a wide range of applications; see~\cite{Kari&Volkov} or \cite{Volkov:2022} for an introduction and overview. Much research on \sa{} per se centers around the question of how the \rl{} of a \san{} relates to its number of states. The results collected in Section~\ref{sec:results} belong to this line of research.

The \emph{\v{C}ern\'y conjecture} asserts that the \rl{} of every \san{} with $n$ states does not exceed $(n-1)^2$. See~\cite[Section~3.1]{Volkov:2022} for the history of the conjecture, which dates back to the 1960s. As of the time of composing this version of the list of results (\today), the \v{C}ern\'y conjecture remains neither confirmed nor refuted\footnote{The late Avraham Trahtman posted several texts on arXiv \cite{Trahtman:2012}--\cite{Trahtman:2022} and ResearchGate~\cite{Trahtman:2022a}--\cite{Trahtman:2024a}, claiming to have proved the \v{C}ern\'y conjecture. The latest of these texts appeared in February 2024. However, in an email to me dated February 10, 2024, Avraham wrote that he had found an error in his proof. He added that he hoped to fix it, but it is unlikely that he succeeded before he passed away on July 17, 2024.}.

Jan \v{C}ern\'{y}~\cite{Cerny:1964} introduced a series of \sa{} $\{\mathrsfs{C}_{n}\}_{n=2,3,\dots}$. The automaton $\mathrsfs{C}_n$ has $n$ states, 2 input letters, and \rl\ $(n-1)^2$. The states of $\mathrsfs{C}_{n}$ are the residues modulo $n$, on which the input letters $a$ and~$b$ act as follows:
\[
0\dt a:=1,\ \ m\dt a:=m\ \text{for $0<m<n$,}\quad m\dt b:=m+1\!\!\pmod{n}.
\]
Fig.~\ref{fig:cerny-n} shows a generic automaton from the \v{C}ern\'{y} series\footnote{We adopt the convention that edges with multiple labels represent bunches of parallel edges. In particular, the edge $0\xrightarrow{a,b}1$ in Fig.~\ref{fig:cerny-n} represents the edges $0\xrightarrow{a}1$ and $0\xrightarrow{b}1$.}.
\begin{figure}[ht]
\begin{center}
\begin{tikzpicture}[scale=0.7,->,>=latex,shorten >=1pt,auto,semithick]
  \tikzstyle{every state}=[minimum size=8mm,inner sep=0pt]

  \node[state] (n0) at (0,-1) {1};
  \node[state] (n1) at (-3,-3) {$0$};
  \node[state] (n2) at (3,-3) {2};
  \node[state] (n3) at (-1.8,-6) {$n{-}1$};
  \node[state] (n4) at (1.8,-6) {3};

  \path (n1) edge node[above left] {$a,b$} (n0)
        (n0) edge node[above] {$b$} (n2)
        (n2) edge node[right] {$b$} (n4)
        (n3) edge node[left] {$b$} (n1);

\path
 (n2) edge[-latex, loop, out = 40, in = 0, distance = 1.5cm] node[above right] {$a$} (n2)
 (n3) edge[-latex, loop, out = -110, in = -150, distance = 1.5cm] node[below left] {$a$} (n3)
 (n4) edge[-latex, loop, out = -20, in = -60, distance = 1.5cm] node[below right] {$a$} (n4)
 (n0) edge[-latex, loop, out = 110, in = 70, distance = 1.5cm] node[above] {$a$} (n0);

  \node at (.1,-6.2) {$\dots$};
\end{tikzpicture}
\end{center}
\caption{The automaton $\mathrsfs{C}_n$}\label{fig:cerny-n}
\end{figure}
\v{C}ern\'{y} has proved that the shortest reset word for $\mathrsfs{C}_{n}$ is the word $(ab^{n-1})^{n-2}a$ of length $(n-1)^2$.

The \emph{\v{C}ern\'{y} function} $\mathfrak{C}(n)$ is defined as the maximum \rl\ of all \sa\ with $n$ states. The above property of the series $\{\mathrsfs{C}_{n}\}$ yields the inequality $\mathfrak{C}(n)\ge(n-1)^2$. The \v{C}ern\'{y} conjecture amounts to the claim that in fact the {equality} $\mathfrak{C}(n)=(n-1)^2$ holds.

As of September 2025, not only the truth of the equality $\mathfrak{C}(n)=(n-1)^2$, but also the existence of any upper bound on $\mathfrak{C}(n)$ that is quadratic in $n$, remain unknown. The best upper bound known to date is the following cubic one, due to Yaroslav Shitov~\cite{Shitov:2019}:
\begin{equation*}
\label{eq:shitov}
\mathfrak{C}(n)\le\left(\dfrac{7}{48}+\dfrac{15625}{798768}\right)n^3+o(n^3),
\end{equation*}
with the leading coefficient close to 0.1654. Shitov's upper bound only slightly improves on the earlier bound, found by Marek Szyku\l{}a~\cite{Szykula:2018}:
\begin{equation*}
\label{eq:szykula}
\mathfrak{C}(n)\le\dfrac{85059n^3+90024n^2+196504n-10648}{511104},
\end{equation*}
with the leading coefficient $\approx 0.1664$. In turn, Szyku\l{}a's bound was a tiny improvement on a result from the 1980s, known as the Pin--Frankl bound:
\begin{equation*}
\label{eq:pin}
\mathfrak{C}(n)\le\dfrac{n^3-n}6,
\end{equation*}
where the leading coefficient is $\frac16=0.1666\dotsc$. See~\cite[Section~3.2]{Volkov:2022} for the history of the Pin--Frankl bound and its proof; the original proof comes from combining results by Jean-\'Eric Pin~\cite{Pin:1983} and P\'eter Frankl~\cite{Frankl:1982}.

Since the \v{C}ern\'{y} conjecture has proved difficult to resolve in general, many researchers have considered its restrictions to certain special classes of automata. If $\mathbf{S}$ is a class of DFAs, the \v{C}ern\'{y} function $\mathfrak{C}_{\mathbf{S}}(n)$ restricted to $\mathbf{S}$ is defined as the maximum \rl\ of all $n$-state \sa\ from $\mathbf{S}$. The results surveyed in Section~\ref{sec:results} aimed to determine, or at least upper-bound, the function $\mathfrak{C}_{\mathbf{S}}(n)$ for various classes $\mathbf{S}$.

\subsection{Methods} Let $\mA=\langle Q,\Sigma\rangle$ be a DFA, $P$ a non-empty subset of $Q$, and $v$ a word over $\Sigma$. The \emph{image} and the \emph{preimage} of $P$ under $v$ are defined as
\[
P\dt v:=\{p\dt v\mid p\in P\}\quad \text{and}\quad Pv^{-1}:=\{q\in Q\mid q\dt v\in P\},
\]
respectively. With this notation, the fact that a word $w\in\Sigma^*$ resets $\mA$ to a state $s\in Q$ can be expressed by either of the equalities $Q\dt w=\{s\}$ or $\{s\}w^{-1}=Q$. These equalities indicate two possible directions in constructing reset words: \emph{compression} (top-down) and \emph{extension} (bottom-up).

The top-down approach starts with the set $Q$ of all states and applies words $u_1,u_2,u_3,\dots$ such that $u_1$ shrinks $Q$, $u_2$ shrinks the image $Q\dt u_1$, $u_3$ shrinks the image $Q\dt u_1u_2$, and so on, until the image $Q\dt u_1u_2\cdots u_m$ becomes a singleton, and hence, $w:=u_1u_2\cdots u_m$ is a reset word. For general synchronizing automata, greedy compression, in which each next word $u_i$ is chosen to be the shortest possible word that strictly shrinks the preceding image, yields reset words whose lengths witness the Pin--Frankl bound.

The bottom-up approach begins with a state $s\in Q$ and seeks words $v_1,v_2,v_3,\dots$ such that the size of $\{s\}v_1^{-1}$ is greater than 1, the size of $(\{s\}v_1^{-1})v_2^{-1}$ is greater than that of $\{s\}v_1^{-1}$, and so on, until the preimage under $v_m$ of $(\dots((\{s\}v_1^{-1})v_2^{-1})\dots)v_{m-1}^{-1}$ is all of~$Q$. Then $w:=v_m\cdots v_2v_1$ is a reset word.

Given a positive constant $\alpha$, a DFA with $n$ states is said to be $\alpha$-\emph{extensible} if for each proper non-singleton subset $P$ of its states, there is a word $v$ of length at most $\alpha n$ such that $|Pv^{-1}|>|P|$. The following observation was stated (without proof) as Proposition 2.3 in \cite{Dubuc:1996}, with a reference to an unpublished manuscript by Savick\'y and Van\v{e}\v{c}ek; for a proof, see \cite{Volkov:2022}, where the result appears as Proposition 3.4. A similar observation was made in \cite[Theorem~2]{Rystsov:1995a}.
\begin{proposition}
\label{prop:extensibility}
Each $\alpha$-extensible DFA with $n>2$ states is synchronizing, and its reset threshold does not exceed $1+\alpha n(n-2)$. In particular, the \v{C}ern\'{y} conjecture holds for $1$-extensible automata.
\end{proposition}

Arguments using induction on the number of states often involve the concepts of subautomata and quotients.

A \emph{subautomaton} of a DFA $\langle Q,\Sigma,\delta\rangle$ is any DFA of the form $\langle S,\Sigma,\tau\rangle$, where non-empty $S\subseteq Q$ satisfies $\delta(s,a)\in S$ for all $s\in S$ and $a\in\Sigma$, and $\tau\colon S\times\Sigma\to S$ is the restriction of $\delta$ to $S\times\Sigma$.

A \emph{congruence} on a DFA $\mathrsfs{A}=\langle Q,\Sigma,\delta\rangle$ is an equivalence relation $\pi$ on the set $Q$ such that, for all $p,q\in Q$ and $a\in\Sigma$, if $(p,q)\in\pi$ then $\bigl(\delta(p,a),\delta(q,a)\bigr)\in\pi$. The \emph{quotient} $\mathrsfs{A}/\pi$ is the DFA $\langle Q/\pi,\Sigma,\delta_\pi\rangle$, where $Q/\pi$ is the set of all $\pi$-classes $[q]_\pi$ for $q\in Q$, and the function $\delta_\pi\colon Q/\pi\times\Sigma\to Q/\pi$ is defined by $\delta_\pi([q]_\pi,a):=[\delta(q,a)]_\pi$.

 A DFA is \emph{an automaton with zero} if it contains a state that is fixed by every letter. The following reduction is sometimes useful.
\begin{proposition}[{\!\!\cite[Lemma 3.3]{Volkov:2022}}]
\label{prop:reduction}
Let $\mathbf{S}$ be any class of automata closed under taking subautomata and quotients, and let $\mathbf{S}^0$ and $\mathbf{S}^{\mathrm{SC}}$ be the classes of automata with zero  from $\mathbf{S}$ and \scn{} automata from $\mathbf{S}$, respectively. Let $f\colon\mathbb{Z}^+\to\mathbb{N}$ be a function such that
\begin{equation}\label{eq:inequality}
f(n)\ge f(n-m+1)+f(m) \text{ for all }\ n\ge m\ge 1.
\end{equation}
If $\mathfrak{C}_{\mathbf{S}^0}(n)\le f(n)$ and $\mathfrak{C}_{\mathbf{S}^{\mathrm{SC}}}(n)\le f(n)$, then $\mathfrak{C}_{\mathbf{S}}(n)\le f(n)$.
\end{proposition}

\section{Annotated list of results}
\label{sec:results}

We list results establishing the validity of the \v{C}ern\'y conjecture when restricted to various classes of DFAs. In addition, we list classes of automata for which the \v{C}ern\'y conjecture has not yet been proved, but a quadratic (in the number of states) upper bound on the reset threshold is known. The list extends that of~\cite[Section~3.5]{Volkov:2022} and follows a similar format. For each class $\mathbf{S}$ of automata, the following information is provided:
\begin{itemize}
\itemsep -1pt
\item \textsl{definition} of the class $\mathbf{S}$,
\item \textsl{upper bound} on the function $\mathfrak{C}_{\mathbf{S}}(n)$,
\item \textsl{method} used to obtain the upper bound,
\item \textsl{lower bound} for the function $\mathfrak{C}_{\mathbf{S}}(n)$,
\item \textsl{comment} (if necessary),
\item \textsl{version stamp} (only for entries that have been substantially updated compared to~\cite[Section~3.5]{Volkov:2022} or newly added)
\end{itemize}
In cases where the upper and lower bounds coincide---that is, where the function $\mathfrak{C}_{\mathbf{S}}(n)$ is known---the class label is marked with a superscript~$\dag$.

When reproducing items from \cite[Section 3.5]{Volkov:2022}, we have silently corrected typographical and translation errors. Furthermore, items referring to results and notions introduced elsewhere in the survey~\cite{Volkov:2022} have been reorganized so as to make the present annotated list reasonably self-contained.

\section*{A. Classes whose function $\mathfrak{C}_{\mathbf{S}}(n)$ is equal or close to $(n-1)^2$}

\noindent\textbf{A1$^\dag$. Circular automata}

\textsl{$\bullet$ Definition:} A DFA is called \emph{circular} if one of its letters acts as a cyclic permutation of the entire set of states.

\textsl{$\bullet$ Upper bound:} $(n-1)^2$; see~\cite{Dubuc:1998}.

\textsl{$\bullet$ Method:} Circular \sa{} are 1-extensible~\cite[Proposition 4.6]{Dubuc:1998}, so Proposition~\ref{prop:extensibility} applies.

\textsl{$\bullet$ Lower bound:} $(n-1)^2$, since the automata $\mC_n$ from the \v{C}ern\'y series are circular.

\textsl{$\bullet$ Comment:} For a circular \san{} with $n$ states, greedy compression yields a reset word of length at most $n^2\ln n$ \cite[Theorem 13]{Gusev&Jungers&Prusa:2018}.

\textsl{$\bullet$ Version stamp:} August 2025 update (comment added).

\medskip

\noindent\textbf{A2. One-cluster automata with a cycle of prime length}

\textsl{$\bullet$ Definition:} A DFA $\langle Q,\Sigma\rangle$ is \emph{one-cluster} for a letter $a\in\Sigma$ if
\[
\bigcap_{q\in Q}\{q\dt a^k\mid k\ge1\}\ne\varnothing.
\]
In this case, the graph representation of the DFA contains exactly one $a$-\emph{cycle}, i.e., a simple cycle all of whose edges are labeled by $a$. Class \textbf{A2} consists of one-cluster automata for some letter $a$ in which the length of the $a$-cycle is prime.

\textsl{$\bullet$ Upper bound:} $(n-1)^2$; see~\cite{Steinberg:2011b}.

\textsl{$\bullet$ Method:} One-cluster \sa{} are 2-extensible. This gives a quadratic upper bound (see item D1), but some extra tricks, making use of the assumption of this item, are needed to reduce the bound to $(n-1)^2$.

\textsl{$\bullet$ Lower bound:} If $n$ is prime, then the tight lower bound is $(n-1)^2$, since the automaton $\mC_n$ from the \v{C}ern\'y series falls into class \textbf{A2} when $n$ is prime. ($\mC_n$ is one-cluster for the letter $b$, and its $b$-cycle has length $n$.)

If $n$ is not prime, then the lower bound is $(n-2)(n-n^{0.525})+2$ for sufficiently large $n$. This bound is witnessed by a series of \sa{} $\mD_{n,p}$, where $n=4,6,8,9,\dots$ is a composite number and $p$ is the greatest prime less than $n$. The states of $\mathrsfs{D}_{n,p}$ are the residues modulo~$n$, on which the input letters $a$~and $b$ act as follows:
\[
m\dt a:=\begin{cases}
0&\text{if }\ m=p-1,\\
n-p&\text{if }\ m=n-1,\\
m+1&\text{if }\ m\ne p-1,n-1;
\end{cases}
\qquad m\dt b:=m+1\!\!\pmod{n}.
\]
The automaton $\mathrsfs{D}_{n,p}$ is one-cluster for $a$, and its $a$-cycle has length $p$, so it lies in \textbf{A2}. For illustration, Fig.~\ref{fig:d10-7} shows the automaton $\mD_{10,7}$.
\begin{figure}[hbt]
\begin{center}
\begin{tikzpicture}[scale=0.8]
	\foreach \ev in {0,1,2,3,4,5,6,7,8,9}
	{
		\node[fill=white, circle, draw=blue, scale=1] at ($({36*(-\ev) + 90}:3cm)$) (\ev) {$\ev$};
		
	}	

	\draw
		(0) edge[-latex, above] node{$a,b$} (1)
		(1) edge[-latex, right] node{$a,b$} (2)
		(2) edge[-latex, right] node{$a,b$} (3)
		(3) edge[-latex, right] node{$a,b$} (4)
		(4) edge[-latex, below] node{$a,b$} (5)
		(5) edge[-latex, below] node{$a,b$} (6)
		(6) edge[-latex, left] node{$b$} (7)
		(7) edge[-latex, left] node{$a,b$} (8)
		(8) edge[-latex, left] node{$a,b$} (9)
		(9) edge[-latex, above] node{$b$} (0)
		
		(6) edge[-latex, left] node{$a$} (0)		
    (9) edge[-latex, above] node{$a$} (3)		
		
	;
\end{tikzpicture}
\end{center}
\caption{The one-cluster automaton $\mD_{10,7}$}\label{fig:d10-7}
\end{figure}

See Appendix A2 for the proof that $\rt(\mathrsfs{D}_{n,p})\ge (n-2)(n-n^{0.525})+2$ for sufficiently large $n$.

\textsl{$\bullet$ Version stamp:} August 2025 update (the lower bound for composite number of states added).

\medskip

\noindent\textbf{A3$^\dag$. Orientable automata}
\nopagebreak

\textsl{$\bullet$ Definition:} A DFA $\langle Q,\Sigma\rangle$ is \emph{orientable}\footnote{The concept goes back to the paper by David Eppstein~\cite{Eppstein:1990}, where such automata were called monotonic. We follow the terminology of~\cite{Ananichev&Volkov:2004}, reserving the term `monotonic' for a narrower class of DFAs; see item C1.} if the states in $Q$ can be arranged in a linear order $q_0<q_1<\dots<q_{n-1}$, with $n=|Q|$, such that for every letter $a\in\Sigma$, the sequence $q_0\dt a,q_1\dt a,\dots,q_{n-1}\dt a$ is \emph{properly oriented}, that is, a cyclic permutation of a nondecreasing sequence.

\textsl{$\bullet$ Upper bound:} $(n-1)^2$, see~\cite{Eppstein:1990}.

\textsl{$\bullet$ Method:} Call a non-empty subset $I\subseteq Q$ an \emph{oriented interval} if for any distinct $p,r\in I$ it contains all states $q\in Q$ such that the sequence $p,q,r$ is properly oriented. (Note that according to this definition, all singletons are intervals.) The crucial property of orientable automata is that for any word $w\in\Sigma^*$ and any oriented interval $I\subseteq Q$, the preimage $Iw^{-1}$ is an oriented interval~\cite[Lemma~1]{Eppstein:1990}. Hence, if $w:=a_1\dots a_\ell$, with $a_1,\dots,a_\ell\in\Sigma$, is a reset word of minimum length for an orientable \san{}, built bottom-up starting from a state $q\in Q$, then all preimages $\{q\}(a_ia_{i+1}\cdots a_\ell)^{-1}$ are distinct non-singleton oriented intervals. A linearly ordered $n$-element set has $(n-1)^2$ distinct non-singleton oriented intervals, whence $\ell\le(n-1)^2$.

\textsl{$\bullet$ Lower bound:} $(n-1)^2$, since the automata $\mC_n$ from the \v{C}ern\'y series are orientable under the standard state ordering $0<1<\dots<{n-1}$.

\textsl{$\bullet$ Comment:} In~\cite[Theorem 2.2]{Ananichev&Volkov:2004}, the upper bound $(n-1)^2$ on \rl{} is extended to weakly orientable \sa{} with $n$ states. A DFA $\langle Q,\Sigma\rangle$ is \emph{weakly orientable} if the states in $Q$ can be arranged in a linear order $q_0<q_1<\dots<q_{n-1}$, with $n=|Q|$, such that for every letter $a\in\Sigma$, one of the sequences $q_0\dt a,q_1\dt a,\dots,q_{n-1}\dt a$ and $q_{n-1}\dt a,q_{n-2}\dt a,\dots,q_0\dt a$ is properly oriented. Yet another, more cumbersome-to-formulate generalization is discussed in item A4.

\textsl{$\bullet$ Version stamp:} August 2025 update (compact definition of orientable automata)

\medskip

\noindent\textbf{A4$^\dag$. Automata respecting the intervals of a directed graph}

\textsl{$\bullet$ Definition:} Let $\Delta$ be a directed graph whose vertex set is the state set of a DFA $\mathrsfs{A}=\langle Q,\Sigma\rangle$. (Warning: in general, $\Delta$ is different from the underlying graph of $\mathrsfs{A}$, although the two graphs share the same vertex set.) For $p,r\in Q$, a path from $p$ to $r$ in $\Delta$ is called \emph{singular} if neither $p$ nor $r$ occurs on this path as an intermediate vertex. Let $[p,r]:=
\varnothing$ if $\Delta$ has no path from $p$ to $r$, and otherwise let
\[
[p,r]:=\{q\in Q \mid q\ \text{lies on a singular path from $p$ to $r$}\}.
\]
The graph $\Delta$ is \emph{dense} if, for all $p,q,r$ in the same \scc{} of $\Delta$, either $q\in[p,r]$ or $q\in[r,p]$.

The DFA $\mathrsfs{A}$ \emph{respects the intervals of the graph} $\Delta$ if the following hold for all $p,r\in Q$ and every $a\in\Sigma$:
\begin{itemize}
\itemsep -1pt
\item[-] if $[p,r]\ne\varnothing$, then $[p\dt a,r\dt a]\ne\varnothing$,
\item[-] if $[p,r]\ne\varnothing$ and $[r,p]\ne\varnothing$, then $[p,r]\dt a\subseteq [p\dt a,r\dt a]$,
\item[-] if $p\dt a=r\dt a$, then at least one of $[p,r]\dt a$ or $[r,p]\dt a$ is a singleton.
\end{itemize}
Class \textbf{A4} consists of strongly connected automata respecting the intervals of some weakly connected dense graph on their state sets.

\textsl{$\bullet$ Upper bound:} $(n-1)^2$, see \cite[Corollary 4.1]{Grech&Kisielewicz:2013}.

\textsl{$\bullet$ Method:} Induction on the number of states combined with a case-by-case analysis.

\textsl{$\bullet$ Lower bound:} $(n-1)^2$, since the automata $\mC_n$ from the \v{C}ern\'y series lie in \textbf{A4}. (The DFA $\mC_n$ is strongly connected and respects the intervals of the cycle $0\to1\to2\to\dots\to n-1\to 0$, which is a dense graph.)

\medskip

\noindent\textbf{A5$^\dag$. 2-junction automata}
\nopagebreak

\textsl{$\bullet$ Definition:} A DFA with the input alphabet $\Sigma$ is a \textup2-\emph{junction automaton} if for every letter $a\in\Sigma$, one of the following conditions holds:
\begin{itemize}
\item[-] all states not fixed by $a$, except for at most two, are fixed by every letter from $\Sigma \setminus \{a\}$, and each of the exceptional states is not fixed by exactly one letter from $\Sigma \setminus \{a\}$,
\item[-] all states not fixed by $a$, except for one, are fixed by every letter from $\Sigma \setminus \{a\}$, and the exceptional state is not fixed by exactly two letters from $\Sigma\setminus\{a\}$.
\end{itemize}

\textsl{$\bullet$ Upper bound:} $(n-1)^2$, see \cite[Theorem 2.1]{Grech&Kisielewicz:2020}.

\textsl{$\bullet$ Method:} Reduction to DFAs in classes \textbf{A1} and \textbf{A6} and an analysis of the remaining cases.

\textsl{$\bullet$ Lower bound:} $(n-1)^2$, as the \v{C}ern\'y automata $\mC_n$ lie in \textbf{A5}.

\medskip

\noindent\textbf{A6. Eulerian automata}

\textsl{$\bullet$ Definition:} A DFA is \emph{Eulerian} if its underlying graph is Eulerian, that is, it contains a cycle that traverses each edge exactly once.

\textsl{$\bullet$ Upper bound:} $n^2-3n+3$, see \cite[Theorem 3]{Kari:2003}.

\textsl{$\bullet$ Method:} Eulerian \sa{} are 1-extensible, moreover, in an Eulerian \san{} $\langle Q,\Sigma\rangle$ with $n$ states, each proper non-singleton subset $P\subset Q$ admits a word $v$ of length at most $n-1$ such that $|Pv^{-1}|>|P|$, see \cite[Lemma 4]{Kari:2003}. The proof on Proposition~\ref{prop:extensibility} then yields a reset word of length at most $1+(n-2)(n-1)=n^2-3n+3$.

\textsl{$\bullet$ Lower bound:} $\lfloor\frac{n^2-3}2\rfloor$, see~\cite{Szykula&Vorel:2016}.

\textsl{$\bullet$ Comments:} The upper bound $n^2-3n+3$ on \rl{} is extended in~\cite[Theorem 3]{Steinberg:2011a} to pseudo-Eulerian \sa{} with $n$ states. A DFA $\langle Q,\Sigma\rangle$ is \emph{pseudo-Eulerian} if the letters of $\Sigma$ can be assigned positive weights summing to 1 such that, for every state $q\in Q$, the total weight of the letters labeling the edges entering $q$ equals 1. (For Eulerian automata, this holds when each letter is assigned the weight $\frac1{|\Sigma|}$.)

The lower bound $\lfloor\frac{n^2-3}2\rfloor$ from~\cite{Szykula&Vorel:2016} is attained by a series of Eulerian automata with four input letters. The automata in this series also witness the tightness of the upper bound $n-1$ on the length of `extending' words from \cite[Lemma 4]{Kari:2003}. Marek Szyku\l{}a and Vojt\v{e}ch Vorel have conjectured that for $n\ge 3$, $\lfloor\frac{n^2-3}2\rfloor$ is also an upper bound for the reset threshold
of $n$-state Eulerian \sa{}~\cite[Conjecture 19]{Szykula&Vorel:2016}. They report that exhaustive searches over small synchronizing Eulerian DFAs verified the bound for the following cases: DFAs with two letters and $\le 11$ states, with four letters and $\le 7$ states, with eight letters and $\le 5$ states, and all DFAs with at most 4 states.

In the class of Eulerian automata with two letters, Pavel Martyugin (unpublished) proposed a series of $n$-state synchronizing automata (with $n$ odd) conjectured to have a reset threshold of $\frac{n^2-5}2$. The exhaustive search experiments in~\cite{Szykula&Vorel:2016} confirmed Martyugin's conjecture for $n = 5, 7, 9, 11$; moreover, for each of these values of $n$, Martyugin's automaton turned out to be the only one with reset threshold $\frac{n^2-5}2$. However, in the general case the conjecture remains unproved and the greatest lower bound for the maximum reset threshold of Eulerian synchronizing automata with two letters and $n$ states (with $n$ odd) currently available in the literature is $\frac{n^2-3n+ 4}{2}$ from~\cite{Gusev:2011}.

\textsl{$\bullet$ Version stamp:} August 2025 update (Szyku\l{}a--Vorel's conjecture added)

\medskip

\noindent\textbf{A7. Automata with a letter of small rank}

\textsl{$\bullet$ Definition:} A DFA $\langle Q,\Sigma\rangle$ is an \emph{automaton with a letter of small rank} if there is a letter $a\in\Sigma$ such $|Q\dt a|\le\sqrt[3]{6|Q|-6}$.

\textsl{$\bullet$ Upper bound:} $(n-1)^2$, see \cite[Corollary 13]{Berlinkov&Szykula:2016}.

\textsl{$\bullet$ Method:} It is a consequence of \cite[Theorem~6]{Berlinkov&Szykula:2016} obtained with an enhanced extension method.

\textsl{$\bullet$ Lower bound:} Unknown.

\textsl{$\bullet$ Comment:} A pioneering result on the \v{C}ern\'y conjecture for automata with a letter of small rank was obtained in \cite[Theorem 1]{Pin:1978b}, under the much stronger condition $|Q\dt a|\le1+\log_2|Q|$ on the `compression rate' of $a$.

\medskip

\noindent\textbf{A8. Automata without involutions}

\textsl{$\bullet$ Definition:} A DFA $\langle Q,\Sigma\rangle$ is an \emph{automaton without involutions} if, for all states $q\in Q$ and words $w\in\Sigma^*$, the equality $q=q\dt w^2$ implies $q=q\dt w$. In algebraic terms, this means that the transition monoid of the automaton has no subgroups of even order.

\textsl{$\bullet$ Upper bound:} $(n-1)^2$, see \cite[Theorem 6]{Trahtman:2008}

\textsl{$\bullet$ Method:} Modified approach from~\cite{Trahtman:2007}, see item B2.

\textsl{$\bullet$ Lower bound:} The only known lower bound is the linear one $n+\left\lfloor\frac{n}2\right\rfloor-2$ that follows from \cite{Ananichev:2010}. For \scn{} \sa{} without involutions, no examples are known with \rl{} greater than or equal to the number of states.

\textsl{$\bullet$ Comment:} The proof in \cite{Trahtman:2008} contains a few unclear points.

\medskip

\noindent\textbf{A9$^\dag$. Automata with \scn{} restricted Rystsov graph}

\textsl{$\bullet$ Definition:} Let $\mathrsfs{A}=\langle Q,\Sigma\rangle$ be a DFA and let a word $w\in\Sigma^*$ be such that $|Q\setminus Q\dt w|=1$. The unique state in $Q\setminus Q\dt w$ is the \emph{excluded state} $\excl(w)$ of $w$, and the unique state $p\in Q\dt w$ with $|pw^{-1}|=2$ is the \emph{duplicate state} $\dupl(w)$ of $w$. The \emph{restricted Rystsov graph} of $\mA$ is the directed graph $R(\mA)$ whose vertices are the states of $\mA$ and whose edges are of the form $\excl(w)\to\dupl(w)$ for all words $w\in\Sigma^*$ of length at most $|Q|$ such that $|Q\setminus Q\dt w|=1$. Class \textbf{A9} consists of automata whose restricted Rystsov graphs are \scn. Such automata are always synchronizing.

\textsl{$\bullet$ Upper bound:} $(n-1)^2$.

\textsl{$\bullet$ Method:} Automata $\langle Q,\Sigma\rangle$ with \scn{} restricted Rystsov graphs have the following strong property, which readily implies the validity of the \v{C}ern\'y conjecture for them: for every non-empty subset $S\subseteq Q$, there exists a word $w\in\Sigma^*$ of length at most $|Q|(|Q|-|S|)$ such that $Q\dt w=S$; see \cite[Lemmas 2 and 3]{Casas&Volkov:2023}.

\textsl{$\bullet$ Lower bound:} The lower bound is $(n-1)^2$, since the \v{C}ern\'y automata $\mC_n$ lie in \textbf{A9}. The restricted Rystsov graph of $\mC_n$ is \scn, as it contains the cycle $0\to1\to2\to\dots\to n-1\to 0$. For $i=0,1\dots,n-1$, the edge $i\to i+1\pmod{n}$ occurs in the graph $R(\mC_n)$ since $i=\excl(ab^{i})$ and $i+1\pmod{n}=\dupl(ab^{i})$.

\textsl{$\bullet$ Comment:} The result is a slight extension of \cite[Theorem 12]{Don:2016}.

\textsl{$\bullet$ Version stamp:} Item added in August 2025.

\medskip

\noindent\textbf{A10$^\dag$. Binary automata with a simple idempotent letter}

\textsl{$\bullet$ Definition:} A DFA is \emph{binary} if its input alphabet consists of two letters. An input letter $a$ of an automaton with state set $Q$ is a \emph{simple idempotent} if $|Q\setminus Q\dt a|=1$ and $q\dt a=q\dt a^2$ for all $q\in Q$.  Class \textbf{A10} consists of binary automata in which one of the two letters is a simple idempotent.

\textsl{$\bullet$ Upper bound:} $(n-1)^2$ (unpublished result by Stefan Hoffmann).

\textsl{$\bullet$ Method:} See Appendix A10 for a proof.

\textsl{$\bullet$ Lower bound:} The lower bound is $(n-1)^2$, since the \v{C}ern\'y automata $\mC_n$ lie in \textbf{A10}.

\textsl{$\bullet$ Version stamp:} Item added in August 2025.

\section*{B. Classes whose function $\mathfrak{C}_{\mathbf{S}}(n)$ is between $n$ and $\frac{n(n-1)}2$}

\noindent\textbf{B1$^\dag$. Automata with zero}

\textsl{$\bullet$ Definition:} A DFA is \emph{an automaton with zero} if it contains a state that is fixed by every letter.

\textsl{$\bullet$ Upper bound:} $\frac{n(n-1)}2$, \cite[Theorem 1]{Rystsov_rus:1977}, see also \cite[Theorem 6.1]{Rystsov:1997}.

\textsl{$\bullet$ Method:} Greedy compression.

\textsl{$\bullet$ Lower bound:} The lower bound is $\frac{n(n-1)}2$ for automata with an unbounded alphabet. It is attained by a series of \sa{} $\{\mathrsfs{R}_{n}\}_{n=2,3,\dots}$ found in~\cite[Theorem 2]{Rystsov_rus:1977}, see also \cite[Theorem 6.1]{Rystsov:1997}. The DFA $\mathrsfs{R}_n$ has $\{0,1,\dots,n-1\}$ as the state set and $\Sigma:=\{a_1,\dots,a_{n-1}\}$ as the input alphabet. The letter $a_1$ fixes all states except state 1, which it maps to 0, and for $2\le i \le n-1$, the letter $a_i$ fixes all states except states $i-1$ and $i$, which it swaps. Fig.~\ref{fig:rystsov-n} shows a generic automaton from the series  $\{\mathrsfs{R}_{n}\}$.
\begin{figure}[htb]
\begin{center}
\begin{tikzpicture}[->,>=latex,shorten >=1pt,auto,node distance=22mm,semithick]
  \tikzstyle{every state}=[minimum size=8mm,inner sep=0pt]

  \node[state] (A) {0};
  \node[state] (B) [right of=A] {1};
  \node[state] (C) [right of=B] {2};
  \node[state] (D) [right of=C] {$3$};
  \node (dots) [right of=D,xshift=-10mm] {$\cdots$};
  \node[state] (E) [right of=dots,xshift=-10mm] {$n{-}2$};
  \node[state] (F) [right of=E] {$n{-}1$};

  \path (B) edge node[below] {$a_1$} (A)

        (C) edge[bend left=20] node[below] {$a_2$} (B)
        (B) edge[bend left=20] node[above] {$a_2$} (C)

        (D) edge[bend left=20] node[below] {$a_3$} (C)
        (C) edge[bend left=20] node[above] {$a_3$} (D)

        (F) edge[bend left=20] node[below] {$a_{n-1}$} (E)
        (E) edge[bend left=20] node[above] {$a_{n-1}$} (F);

  \path (A) edge[loop above,distance = 1cm] node[align=center] {\footnotesize $\Sigma$} (A)
        (B) edge[loop above,distance = 1cm] node[align=center] {\footnotesize $\Sigma{\setminus}\{a_1,a_2\}$} (B)
        (C) edge[loop above,distance = 1cm] node[align=center] {\footnotesize $\Sigma{\setminus}\{a_2,a_3\}$} (C)
        (D) edge[loop above,distance = 1cm] node[align=center] {\footnotesize $\Sigma{\setminus}\{a_3,a_4\}$} (D)
        (E) edge[loop above,distance = 1cm] node[align=center] {\footnotesize $\Sigma{\setminus}\{a_{n{-}2},a_{n{-}1}\}$} (E)
        (F) edge[loop above,distance = 1cm] node[align=center] {\footnotesize $\Sigma{\setminus}\{a_{n{-}1}\}$} (F);
\end{tikzpicture}\caption{The automaton $\mathrsfs{R}_n$}\label{fig:rystsov-n}
\end{center}
\end{figure}

For automata with two input letters, the best currently known lower bound for the maximum reset threshold of $n$-state synchronizing automata with zero is $\frac{1}{4}n^2+2n-9$. This bound is attained for $n\equiv4\pmod{12}$, starting from $n=16$, by a series of automata presented in~\cite{Ananichev&Vorel:2019}. Other series of $n$-state synchronizing automata in \textbf{B1} with a fixed alphabet size and reset threshold of order $\frac{1}{4}n^2+O(n)$ are exhibited in~\cite{Martugin:2009,Pribavkina:2011}. The automata in~\cite{Martugin:2009} have two letters, $n\ge 8$, and reset threshold  $\lceil\frac{1}{4}n^2+\frac32n-4\rceil$, while those in \cite{Pribavkina:2011} have any fixed number $k$ of letters, even $n\ge 2k$, and reset threshold $\frac{1}{4}n^2+\frac12n-1$.

\medskip

\noindent\textbf{B2. Aperiodic automata}
\nopagebreak

\textsl{$\bullet$ Definition:} A DFA $\langle Q,\Sigma\rangle$ is \emph{aperiodic} if, for all states $q\in Q$, words $w\in\Sigma^*$, and positive integers $k$, the equality $q=q\dt w^k$ implies $q=q\dt w$. In algebraic terms, this means that the transition monoid of the automaton has no nontrivial subgroups.

\textsl{$\bullet$ Upper bound:} $\frac{n(n-1)}2$, see \cite[Theorem 10]{Trahtman:2007}.

\textsl{$\bullet$ Method:} The class of aperiodic automata is closed under taking subautomata and quotients, and the function $\frac{n(n-1)}2$ satisfies inequality~\eqref{eq:inequality}. By Proposition~\ref{prop:reduction}, it suffices to verify the upper bound $\frac{n(n-1)}2$ for strongly connected synchronizing aperiodic automata and synchronizing aperiodic automata with zero. In the second case the bound holds for every $n$-state \san{} with zero (see B1). A strongly connected aperiodic DFA $\langle Q,\Sigma\rangle$ is always synchronizing \cite[Theorem 9]{Trahtman:2007} and admits a nontrivial partial order $\le$ stable under the action of input letters: if $p\le r$ for some $p,r\in Q$, then $p\dt a\le r\dt a$ for each letter $a\in\Sigma$. Therefore, if a state $p$ is minimal and a state $r$ is maximal with respect to $\le$, then every word sending $p$ to $r$ sends every state $q\in Q$ with $p\le q$ to $r$. Similarly, every word sending $r$ to $p$ sends every state $q\in Q$ with $q\le r$ to $p$. In any strongly connected automaton with $n$ states, for any pair $(q,q')$ of states there is a word of length at most $n-1$ that sends $q$ to $q'$. This makes it possible to reset any strongly connected aperiodic DFA with $n$ states by a word of length $\le\frac{n(n-1)}2$ composed from words of length $\le n-1$ sending maximal states to minimal ones or vice versa, because either the number of maximal states or the number of minimal states does not exceed $\frac{n}2$. A precise implementation of this idea requires induction on the number of states.

\textsl{$\bullet$ Lower bound:} Only the linear lower bound $n+\left\lfloor\frac{n}2\right\rfloor-2$ is currently known; it follows from \cite{Ananichev:2010}. No examples of strongly connected aperiodic $n$-state automata are available with reset threshold higher than $n-1$, and it is conjectured that the reset threshold for such automata does not exceed $n-1$. This conjecture is verified confirmed by an exhaustive search over strongly
connected aperiodic automata with two letters and at most 11 states and with three letters and at most 7 states, see~\cite[Section 7.4.4]{Szykula:2014}.

\textsl{$\bullet$ Comment:} In \cite{Volkov:2009}, the upper bound for the reset threshold of strongly connected aperiodic $n$-state automata was improved to $\left\lfloor\frac{n(n+1)}6\right\rfloor$.

\medskip

\noindent\textbf{B$3^\dag$. Automata with transition monoids in $\mathbf{EDS}$}
\nopagebreak

\textsl{$\bullet$ Definition:} $\mathbf{DS}$ is the class of all finite monoids $M$ such that, for all $x,y,z\in M$, the following implication is satisfied:\footnote{An equivalent and more standard definition in terms of semigroup theory is that $\mathbf{DS}$ is the class of all finite monoids whose regular $\mathcal{D}$-classes are subsemigroups.}
\[
\text{if }\ MxM=MyM=MzM=Mx^2M,\ \text{ then }\ MxM=MyzM.
\]
An \emph{idempotent} is an element $e$ of a monoid $M$ such that $e^2=e$. The class $\mathbf{EDS}$ consists of all finite monoids $M$ such that the submonoid generated by all idempotents in $M$ belongs to $\mathbf{DS}$. The class \textbf{B3} comprises automata whose transition monoids belong to $\mathbf{EDS}$.

\textsl{$\bullet$ Upper bound:} $\frac{n(n-1)}2$, see \cite[Corollary 4.3]{Almeida&Steinberg:2009}.

\textsl{$\bullet$ Method:} The argument employs information about the linear representations of monoids in $\mathbf{EDS}$ given by the classical theory of semigroup representations (the Munn--Ponizovsky theory).

\textsl{$\bullet$ Lower bound:} The lower bound is $\frac{n(n-1)}{2}$ for automata with an unbounded alphabet, since the automata $\mR_n$ (see item B1) belong to \textbf{B3}. Almost nothing is known about lower bounds for the maximum reset threshold for \sa{} in \textbf{B3} with a fixed alphabet size, or for strongly connected \sa{} in this class.

\medskip

\noindent\textbf{B4. Decoders of finite maximal prefix codes}

\textsl{$\bullet$ Definition:} A \emph{proper prefix} of a non-empty word $w$ is any word $u$ such that $w=uv$ for some non-empty word $v$. A set $X$ of non-empty words over an alphabet $\Sigma$ is a \emph{prefix code over $\Sigma$} if no word in $X$ is a proper prefix of another word in $X$. A prefix code is \emph{maximal} if it is not contained in any other prefix code over the same alphabet.

The \emph{decoder of a finite maximal prefix code} $X$ over $\Sigma$ is the DFA $\langle Q,\Sigma\rangle$ such that $Q$ is the set of proper prefixes of words in $X$ and, for all $q\in Q$ and $a\in \Sigma$,
\[
q\dt a:=\begin{cases} qa & \text{if $qa$ is a proper prefix of a word in $X$},\\
\varepsilon & \text{if $qa \in X$}.\end{cases}
\]

\textsl{$\bullet$ Upper bound:} $O(n\log^3n)$. In detail, letting $r:=\lceil\log_{|\Sigma|}n\rceil$, the upper bound is $2+(n+r-1)\frac{r^3-r}6$ for $r\ge 4$ and $2+(n+r-1)(r-1)^2$ for $r\le 3$,  see \cite[Corollary 17]{Berlinkov&Szykula:2016}.

\textsl{$\bullet$ Method:} A combination of greedy compression and greedy extension.

\textsl{$\bullet$ Lower bound:} The lower bound is $2n-5$ for even $n$ and $2n-7$ for odd $n$ if $|\Sigma|=2$, see~\cite{Biskup&Plandowski:2009b}. For $|\Sigma|\ge3$, the lower bound $2\lceil\frac{n}{|\Sigma|+1}\rceil$ is obtained in \cite[Theorem 19]{Berlinkov&Szykula:2016}.

\textsl{$\bullet$ Version stamp:} August 2025 update (details of the upper bound added)

\medskip

\noindent\textbf{B5. Weakly monotonic automata}
\nopagebreak

\textsl{$\bullet$ Definition:} A DFA $\langle Q,\Sigma\rangle$ is \emph{weakly monotonic} if the state set $Q$ admits a linear order such that for each letter $a\in\Sigma$, the function $q\mapsto q\dt a$ is either non-decreasing or non-increasing.

\textsl{$\bullet$ Upper bound:} $\frac{n(n-1)}2$ in general; $2n-3$ if there is a letter $c$ such that the function $q\mapsto q\dt c$ is a non-increasing bijection, see~\cite[Proposition~3.2]{Ananichev&Volkov:2004}.

\textsl{$\bullet$ Method:} Call a non-empty subset $I\subseteq Q$ an \emph{interval} if for any $p,r\in I$ it contains all states $q\in Q$ such that the sequence $p\le q\le r$. In a weakly monotonic automaton, for any word $w\in\Sigma^*$ and any interval $I\subseteq Q$, the preimage $Iw^{-1}$ is an interval.  Hence, if $w:=a_1\dots a_\ell$, with $a_1,\dots,a_\ell\in\Sigma$, is a reset word of minimum length for a weakly monotonic \san{}, built bottom-up starting from a state $q\in Q$, then all preimages $\{q\}(a_ia_{i+1}\cdots a_\ell)^{-1}$ are distinct non-singleton intervals. A linearly ordered $n$-element set has $\frac{n(n-1)}2$ distinct non-singleton intervals, whence $\ell\le\frac{n(n-1)}2$.

If there is a letter that acts as a non-increasing bijection, then a reduction to monotonic DFAs (see item C1) is employed.

\textsl{$\bullet$ Lower bound:} $2n-3$ for $n\equiv3\pmod{4}$, see~\cite[Proposition~3.1]{Ananichev&Volkov:2004}.

\medskip

\noindent\textbf{B$6^\dag$. 0-monotonic automata}
\nopagebreak

\textsl{$\bullet$ Definition:} A DFA $\langle Q,\Sigma\rangle$ with zero  0 is 0-\emph{monotonic}  if the set $Q\setminus\{0\}$ admits a linear order $\le$ such that for each letter $a\in\Sigma$ and all $p,r\in Q\setminus\{0\}$, the relations $p\le r$ and $p\dt a,r\dt a\ne 0$ imply $p\dt a\le r\dt a$.

\textsl{$\bullet$ Upper bound:} $n+\left\lfloor\frac{n}2\right\rfloor-2$, see.~\cite[Theorem~1]{Ananichev:2010}.

\textsl{$\bullet$ Method:} Reduction to some properties of monotonic DFAs, see item C1.

\textsl{$\bullet$ Lower bound:} $n+\left\lfloor\frac{n}2\right\rfloor-2$, see~\cite[Theorem~2]{Ananichev:2010}.

\medskip

\noindent\textbf{B7. Finitely generated automata}

\textsl{$\bullet$ Definition:} A DFA $\mathrsfs{A}$ with input alphabet $\Sigma$ is \emph{finitely generated} if its set of reset words can be represented as $\Sigma^*W\Sigma^*$ for some finite subset $W\subset\Sigma^*$. In algebraic terms, this means that the set of all reset words of $\mathrsfs{A}$ is a finitely generated ideal of the monoid $\Sigma^*$.

\textsl{$\bullet$ Upper bound:} $3n-5$, see~\cite[Theorem~4]{Pribavkina&Rodaro:2011}.

\textsl{$\bullet$ Method:} The argument employs the characterization of finitely generated automata from \cite[Theorem~1]{Pribavkina&Rodaro:2011} and an upper bound from \cite[Proposition 5]{Pin:1978b}.

\textsl{$\bullet$ Lower bound:} $n-1$, see the discussion of Example 1 in \cite{Pribavkina&Rodaro:2011}.

\section*{C. Classes whose function $\mathfrak{C}_{\mathbf{S}}(n)$ is $n-1$}

\noindent\textbf{C$1^\dag$. Monotonic automata}

\textsl{$\bullet$ Definition:} A DFA $\langle Q,\Sigma\rangle$ is \emph{monotonic} if the state set $Q$ admits a linear order such that for each letter $a\in\Sigma$, the function $q\mapsto q\dt a$ is non-decreasing.

\textsl{$\bullet$ Upper bound:} $n-1$, see~\cite{Ananichev&Volkov:2004a}.

\textsl{$\bullet$ Method:} Induction on the number of states.

\textsl{$\bullet$ Lower bound:} The lower bound $n-1$ is attained by monotonic automata $\mathrsfs{M}_n:=\langle\{0,1,2,\dots,n-1\},\{a\}\rangle$ such that $i\dt a:=i-1$ if $i>0$ and $0\dt a=0$.

\textsl{$\bullet$ Comment:} As far as can be judged from the English abstract of the paper~\cite{Cui&He&Sun:2019}, published in Chinese, it extends the upper bound $n-1$ to those $n$-state \sa{} $\langle Q,\Sigma\rangle$ for which the state set $Q$ admits a partial order~$\le$ having both a greatest and a least element, and such that for any letter~$a\in\Sigma$ and all $p,r\in Q$, the relation $p\le r$ implies $p\dt a\le r\dt a$.

\medskip

\noindent\textbf{C$2^\dag$. Generalized monotonic automata}

\textsl{$\bullet$ Definition:} A DFA $\mathrsfs{A}=\langle Q,\Sigma\rangle$ is $\rho$-\emph{monotonic}, where $\rho$ is a congruence on $\mathrsfs{A}$, if its state set $Q$ admits a partial order~$\le$ such that two states are
comparable with respect to $\le$ if and only if they belong to the same $\rho$-class and for any letter~$a\in\Sigma$ and all $p,r\in Q$, the relation $p\le r$ implies $p\dt a\le r\dt a$. A DFA $\mathrsfs{A}$ is
\emph{generalized monotonic} if it admits a chain of congruences
\[
\rho_0\subset\rho_1\subset\dots\subset\rho_\ell
\]
such that $\rho_0$ is the equality relation, $\rho_\ell$ is the universal relation on $Q$, and for each $i=1,\dots,\ell$, the quotient $\mathrsfs{A}/\rho_{i-1}$ is $\rho_i/\rho_{i-1}$-monotonic.

\textsl{$\bullet$ Upper bound:} $n-1$, see~\cite[Theorem 1.2]{Ananichev&Volkov:2005}.

\textsl{$\bullet$ Method:} Induction on the number of states.

\textsl{$\bullet$ Lower bound:} $n-1$ since \textbf{C2} contains \textbf{C1}, and hence, the DFAs $\mathrsfs{M}_n$.

\medskip

\noindent\textbf{C$3^\dag$. Automata with transition monoids in $\mathbf{DS}$}

\textsl{$\bullet$ Definition:} Class \textbf{C3} consists of the automata whose transition monoids belong to $\mathbf{DS}$; see item B3 for the definition of $\mathbf{DS}$.

\textsl{$\bullet$ Upper bound:} $n-1$, see \cite[Theorem 2.6]{Almeida&Steinberg:2009}.

\textsl{$\bullet$ Method:} Employing information about the linear representations of monoids in \textbf{DS} obtained in~\cite{Almeida&Margolis&Steinberg&Volkov:2009}.

\textsl{$\bullet$ Lower bound:} $n-1$, since the automata $\mathrsfs{M}_n$ introduced in item C1 belong to class \textbf{C3}.

\textsl{$\bullet$ Comment:} Class \textbf{C3} contains several classes of automata, for which the upper bound $n-1$ was established earlier in~\cite{Rystsov_rus:1994,Rystsov:1996,Rystsov:1997}. It also includes the class of the so-called weakly cyclic automata for which the same bound was proved in~\cite{Ryzhikov:2019}. In terms of transition monoids, \cite{Rystsov_rus:1994,Rystsov:1996,Rystsov:1997} dealt with automata whose transition monoids are commutative or close to commutative, whereas~\cite{Ryzhikov:2019} studied automata whose transition monoids are $\mathcal{R}$-trivial.\footnote{A monoid $M$ is $\mathcal{R}$-\emph{trivial} if, for all $x, y\in M$, the equality $xM = yM$ implies $x=y$.} It is known that both commutative and $\mathcal{R}$-trivial finite monoids belong to $\mathbf{DS}$.

See also the comment for the next item.

\medskip

\noindent\textbf{C4$^\dag$. Binary automata with idempotent letters}
\nopagebreak

\textsl{$\bullet$ Definition:} A DFA $\langle Q,\Sigma\rangle$ has \emph{idempotent letters} if $q\dt a=q\dt a^2$ for all $q\in Q$ and $a\in\Sigma$. Class \textbf{C4} consists of DFAs with two idempotent letters.

\textsl{$\bullet$ Upper bound:} $n-1$, see \cite[Proposition 5]{Volkov:2019}.

\textsl{$\bullet$ Method:} The class of binary automata with idempotent letters is closed under taking subautomata and quotients, and the function $n-1$ satisfies inequality~\eqref{eq:inequality}. By Proposition~\ref{prop:reduction}, it suffices to verify the upper bound $n-1$ for \sa{} in \textbf{C4} that either are strongly connected or have zero. A strongly connected synchronizing binary automaton with idempotent letters has at most two states. Induction on the number of states is employed for synchronizing automata with zero and two idempotent letters.

\textsl{$\bullet$ Lower bound:} The lower bound $n-1$ is attained by a series of \sa{} $\langle\{1,2,\dots,n\},\{a,b\}\rangle$, whose letters act as follows:
\[
i\dt a:=\begin{cases}
i &\text{if $i$ is odd or $i=n$},\\
i{+}1 &\text{if $i<n$ is even};
\end{cases}\quad
i\dt b:=\begin{cases}
i &\text{if $i$ is even or $i=n$},\\
i{+}1 &\text{if $i<n$ is odd}.
\end{cases}
\]

\textsl{$\bullet$ Comment:} The classification of finite semigroups generated by two idempotents obtained in \cite[Section III-3]{Benzaken&Mayr:1975} shows that the transition monoid of every DFA with two idempotent letters belongs to the class \textbf{C3}. Therefore, the class \textbf{C4} is a subclass of \textbf{C3}, which yields an alternative proof of the upper bound $n-1$.

The reset threshold of an $n$-state \san{} with three  idempotent letters can be as large as $\lfloor\frac{n^2}2\rfloor-2n+2$, see \cite[\S3]{Don&Zantema:2019} or \cite[Corollary 3]{Volkov:2019}.

\textsl{$\bullet$ Version stamp:} December 2025 update (comment that \textbf{C4} is a subclass of \textbf{C3} added)

\medskip

\noindent\textbf{C5$^\dag$. Media}

\textsl{$\bullet$ Definition:} In a DFA $\langle Q,\Sigma\rangle$, two letters $a,b\in\Sigma$ are \emph{reverses} of each other if, for any distinct states $p,r\in Q$, one has $p\dt a=r$ if and only if $r\dt b=p$. A word in $\Sigma^*$ is \emph{consistent} if it does not contain two letters that are reverses of each other. A word $w\in\Sigma^*$ is \emph{vacuous} if each letter $a\in\Sigma$ occurs in $w$ the same number of times as does the reverse of $a$. A word $a_1\cdots a_\ell$, where $a_1,\dots,a_\ell\in\Sigma$, is \emph{stepwise effective for a state $q\in Q$} if $q\dt a_1\ne q$ и $q\dt a_1\cdots a_i\ne q\dt a_1\cdots a_{i-1}$ for all $i=2,\dots,\ell$. A \emph{medium} is a DFA $\langle Q,\Sigma\rangle$ satisfying the following axioms:
\begin{itemize}
\itemsep -1pt
\item[-] every letter has a unique reverse,
\item[-] for any distinct states $p,r\in Q$, there exists a consistent word $w\in\Sigma^*$ such that $p\dt w=r$,
\item[-] if a word $w\in\Sigma^*$ is stepwise effective for a state $q\in Q$, then $q\dt w=q$ if and only if $w$ is vacuous,
\item[-] if $p\dt u=r\dt v$, where $u$ and $v$ are consistent words such that $u$ is stepwise effective for $p$ and $v$ is stepwise effective for $r$, then the word $uv$ is consistent.
\end{itemize}

\textsl{$\bullet$ Upper bound:} $n-1$, see \cite[Theorem 1]{Eppstein&Falmagne:2008}.

\textsl{$\bullet$ Method:} Reduction to a search for a shortest path in an acyclic auxiliary graph on the vertex set $Q$.

\textsl{$\bullet$ Lower bound:} The lower bound $n-1$ is claimed in \cite{Eppstein&Falmagne:2008}, although an example of a medium with reset threshold $n-1$ is is provided only for $n=6$.

\medskip

\noindent\textbf{C6$^\dag$. Strongly semisimple automata}

\textsl{$\bullet$ Definition:} A DFA is $\mA$ is \emph{strongly semisimple} if a word resets $\mA$ whenever so does any of its powers.

\textsl{$\bullet$ Upper bound:} $n-1$, see \cite[Theorem 21]{Almeida&Rodaro:2016}.

\textsl{$\bullet$ Method:} Application of the classical structure theory of semisimple rings (the Wedderburn--Artin theory).

\textsl{$\bullet$ Lower bound:} The lower bound $n-1$ is attained by a series of \sa{} $\langle\{1,2,\dots,n\},\{a_1,\dots,a_{n-1}\}\rangle$, in which, for all $i=1,\dots,n-1$, the letter $a_i$ sends state $i$ to $i+1$ and fixes all other states.

\textsl{$\bullet$ Version stamp:} Item added in August 2025.

\medskip

\noindent\textbf{C7$^\dag$. Automata with simple idempotent letters}
\nopagebreak

\textsl{$\bullet$ Definition:}  See item A10 for the definition of a simple idempotent letter. Class \textbf{C7} consists of DFAs all of whose letters are simple idempotents.

\textsl{$\bullet$ Upper bound:} $n-1$. See \cite[Theorem 1]{Rystsov:2022}, where the result is stated for aperiodic automata in \textbf{C7}. However, the argument in~\cite{Rystsov:2022}, outlined in the next paragraph, does not use the aperiodicity assumption.

\textsl{$\bullet$ Method:} Compression. In detail, let $\mA=\langle Q,\Sigma\rangle$ be a DFA with simple idempotent letters that can be reset to a state $q_0$. For each $q\in Q\setminus\{q_0\}$, fix a shortest path from $q$ to $q_0$ in the graph representing $\mA$. The length of this path is the \emph{height} of $q$. Let $w_0:=\varepsilon$. If $i\in\{1,\dots,n-1\}$ and the word $w_{i-1}$ has already been constructed, take a state $q\in Q\dt w_{i-1}\setminus\{q_0\}$ of maximum height and let $w_i:=w_{i-1}a$ where $a$ is the label of the first edge of the fixed path from $q$ to $q_0$. Then $w_{n-1}$ is a word of length $n-1$ that resets $\mA$ to $q_0$.

\textsl{$\bullet$ Lower bound:} The lower bound $n-1$ is attained by every $n$-state \san{} $\langle Q,\Sigma\rangle$ with simple idempotent letters because, for every non-singleton subset $P\subseteq Q$ and each letter $a\in\Sigma$, one has $|P\dt a|\ge |P|-1$, whence any word $w\in\Sigma^*$ such that $|Q\dt w|=1$ must have at least $n-1$ letters.

\textsl{$\bullet$ Version stamp:} Item added in August 2025.

\section*{D. Classes with a quadratic upper bound on the function $\mathfrak{C}_{\mathbf{S}}(n)$ }

\noindent\textbf{D1. One-cluster automata}
\nopagebreak

\textsl{$\bullet$ Definition:} See item A2.

\textsl{$\bullet$ Upper bound:} $2n^{2}-4n+1-2(n-1)\ln\frac{n}{2}$, see~\cite[Corollary 1]{Carpi&D'Alessandro:2013}.

\textsl{$\bullet$ Method:} One-cluster \sa{} are 2-extensible. Proposition~\ref{prop:extensibility} gives an upper bound of the form $2n^2+o(n^2)$. The $o(n^2)$ term was improved in several papers, see \cite{Beal&Perrin:2009,Carpi&D'Alessandro:2009,Carpi&D'Alessandro:2009a,Beal&Berlinkov&Perrin:2011,Steinberg:2011a,Steinberg:2011b}, and the asymptotically best to date expression was found in
\cite{Carpi&D'Alessandro:2013}.

\textsl{$\bullet$ Lower bound:} $(n-1)^2$, since the automata $\mC_n$ from the \v{C}ern\'y series are one-cluster for the letter $b$.

\textsl{$\bullet$ Comment:} A DFA is \emph{quasi-one-cluster of degree $d$} if it has a letter~$a$ such that the total length of all but one of the simple cycles labeled by $a$ does not exceed $d$. One-cluster
automata are the quasi-one-cluster automata of degree 0; thus, the parameter $d$ measures the ‘deviation’ from the one-cluster structure. The reset threshold of an $n$-state quasi-one-cluster synchronizing DFA of degree $d$ does not exceed $2^{d}(n-d+1)(2n-d-2)$ \cite[Corollary 11]{Berlinkov:2013}.

\medskip

\noindent\textbf{D2. Completely reachable automata}

\textsl{$\bullet$ Definition:} A DFA $\langle Q,\Sigma\rangle$ is \emph{completely reachable} if, for every non-empty subset $P\subseteq Q$, there exists a word $w\in\Sigma^*$ such that $Q\dt w=P$.

\textsl{$\bullet$ Upper bound:} $2n^2-n\ln n - 4n + 2$ for $n\ge 3$, see~\cite[Corollary 31]{Ferens&Szykula:2023}; $2n^2-n\ln n - 4n$ for $n\ge 6$, see Corollary 6.7 in \cite[Version 5]{Ferens&Szykula:2022}.

\textsl{$\bullet$ Method:} Completely reachable automata are 2-extensible; more precisely, for a completely reachable automaton $\langle Q,\Sigma\rangle$ and a non-empty proper subset $P\subset Q$, there exists a word $v\in\Sigma^*$ of length at most $2n - \lceil\frac{n}{n-|P|}\rceil$ such that $|Pv^{-1}|>|P|$, see \cite[Lemma 29]{Ferens&Szykula:2023}.

\textsl{$\bullet$ Lower bound:}  $(n-1)^2$, since the automata $\mC_n$ from the \v{C}ern\'y series are completely reachable \cite[Proposition 2]{Maslennikova:2014}.

\textsl{$\bullet$ Comment:} Class \textbf{D2} includes DFAs of two types for which quadratic upper bounds on \rl{} were established before:

(a) automata whose transition monoid contains all transformations of the state set (see \cite[Section~3.5, item D2]{Volkov:2022}), and

(b) \scn{} \sa{} in which each letter either acts as a permutation of the state set or fixes all states except one (see \cite[Section 3.5, item D3]{Volkov:2022}; for their complete reachability, see \cite[Theorem 3.6]{Hoffmann:2023}).

The current upper bound on the \rl{} of completely reachable automata improves on both the bound $2n^{2}-6n+5$ from \cite[Theorem 7]{Gonze:2019} for the DFAs of type (a) and the bound $2(n-1)^2$ from \cite[Theorem 4]{Rystsov:2000} for those of type (b).

\textsl{$\bullet$ Version stamp:} Item added in August 2025.

\medskip

\noindent\textbf{D3. Quasi-Eulerian automata}

\textsl{$\bullet$ Definition:} A DFA $\langle Q,\Sigma\rangle$ is \emph{quasi-Eulerian of degree $d$} if it satisfies the following two conditions:
\begin{itemize}
\itemsep -1pt
\item[-] there exists a subset $E_d\subset Q$ with $|E_d|=|Q|-d$ such that exactly one state $s\in E_d$ may have incoming edges from the set $Q\setminus E_d$;
\item[-] the letters in $\Sigma$ can be assigned positive weights such that their total weight equals 1, and for every state $r\in E_d\setminus\{s\}$ the total weight of the labels of all edges entering $r$ equals 1.
\end{itemize}

\textsl{$\bullet$ Upper bound:} $2^d(n-d+1)(n-1)$, see~\cite[Theorem 8]{Berlinkov:2013}.

\textsl{$\bullet$ Method:} Enhanced extension method.

\textsl{$\bullet$ Lower bound:} $(n-1)^2$, since the automata $\mC_n$ from the \v{C}ern\'y series are quasi-Eulerian of degree 1, with the set $\{1,2,\dots,n-1\}$ as $E_1$ and the state~1 as $s$, see Fig.~\ref{fig:cerny-n}. The weights assigned to the letters $a$ and $b$ may be chosen as any two positive numbers summing to 1.

\medskip

\noindent\textbf{D4. Regular automata}

\textsl{$\bullet$ Definition:} A DFA $\langle Q,\Sigma\rangle$ is \emph{regular} if there exists a set of words $W\subset\Sigma^*$ containing the empty word and satisfying the following two conditions:
\begin{itemize}
\itemsep -1pt
\item[-] the length of any word in $W$ is less than $|Q|$;
\item[-] there is an integer $m\ge 1$ such that for any two states $p,r\in Q$, there are exactly $m$ words $W$ that send $p$ to $r$.
\end{itemize}

\textsl{$\bullet$ Upper bound:} $2(n-1)^2$, see~\cite[Theorem 7]{Rystsov:1995} or~\cite[Theorem 7]{Rystsov:1995a}.

\textsl{$\bullet$ Method:} Regular \sa{} are 2-extensible, whence Proposition~\ref{prop:extensibility} applies.

\textsl{$\bullet$ Lower bound:} $(n-1)^2$,  since the automata $\mC_n$ from the \v{C}ern\'y series are regular, with $m=1$ and the set of words $\{\varepsilon,b,b^2,\dots,b^{n-1}\}$ as $W$.

\textsl{$\bullet$ Comment:} Similar concepts were introduced in~\cite{Carpi&D'Alessandro:2009a,Carpi&D'Alessandro:2013}. In~\cite{Carpi&D'Alessandro:2009a}, a DFA $\mathrsfs{A}=\langle Q,\Sigma\rangle$ with $|Q|=n$ is called \emph{strongly transitive} if there exist words $w_1,\dots,w_n\in\Sigma^*$ such that, for all $p,r\in Q$, exactly one of the words $w_i$ sends $p$ to $r$. If such an automaton $\mA$ is synchronizing, then $\rt(\mA)\le (n-2)(n+\max\{|w_1|,\dots,|w_n|\}-1)+1$; see~\cite[Theorem 2]{Carpi&D'Alessandro:2009a}.

In~\cite{Carpi&D'Alessandro:2013} a DFA $\mathrsfs{A}=\langle Q,\Sigma\rangle$ with $|Q|=n$ is called \emph{locally strongly transitive} if there exist words $w_1,\dots,w_k\in\Sigma^*$ and distinct states  $q_1,\dots,q_k\in Q$ such that $\{q\dt w_1,\dots,q\dt w_k\}=\{q_1,\dots,q_k\}$ for each $q\in Q$.  If such an automaton $\mA$ is synchronizing, then
\[
\rt(\mA)\le (k-1)(n+\max\{|w_1|,\dots,|w_n|\}+1)-2 k \ln \frac{k+1}{2}+\min\{|w_1|,\dots,|w_n|\};
\]
see~\cite[Proposition 5]{Carpi&D'Alessandro:2013}.

\medskip

\noindent\textbf{D5. Automata with coinciding cycles}
\nopagebreak

\textsl{$\bullet$ Definition:} A DFA $\langle Q,\Sigma\rangle$ is an \emph{automaton with coinciding cycles} if there exists a set of letters $\Pi\subset\Sigma$ with $|\Pi|>1$ satisfying the following three conditions:
\begin{itemize}
\itemsep -1pt
\item[-] each letter $a\in\Pi$ acts as a cyclic permutation on a non-singleton subset $S_a\subseteq Q$ and fixes all states in $Q\setminus S_a$;
\item[-] if $a,b\in \Pi$ are such that $k:=|S_a\cap S_b|>0$, then for some states $q_a,q_b\in S_a\cap S_b$, one has
\[
S_a\cap S_b=\{q_a,q_a\dt a,\dots,q_a\dt a^{k-1}\}=\{q_b,q_b\dt b,\dots,q_b\dt b^{k-1}\};
\]
\item[-] the DFA $\langle Q,\Pi\rangle$ is \scn.
\end{itemize}

\textsl{$\bullet$ Upper bound:} $6n^2-11n-1$, see \cite[Theorem 2]{Ruszil:2023}.

\textsl{$\bullet$ Method:} Compression relying on the following property: if $\mA=\langle Q,\Sigma\rangle$ is an automaton in \textbf{D5} and $\Pi$ is its distinguished set of letters, then for all states $p,q,r,s\in Q$ with $p\ne q$ and $r\ne s$, there exists a word $v\in\Pi^*$ of length at most $6n$ such that $\{p,q\}\dt w=\{r,s\}$; see~\cite[Lemma 5]{Ruszil:2023}. If $\mA$ is synchronizing, then there are two distinct states $r,s\in Q$ and a letter $c\in\Sigma$ such that $r\dt c=s\dt c$. Therefore, to shrink any non-singleton subset $P\subseteq Q$, one can first apply a shortest word $v\in\Pi^*$ such that $\{p,q\}\dt v=\{r,s\}$ for some distinct states $p,q\in P$ and then apply the letter $c$. This gives a reset word $w=cv_1cv_2\cdots v_mc$ with $m\le n-2$ and $|v_i|\le 6n$, whence $|w|\le 1+(6n+1)(n-2)=6n^2-11n-1$.

\textsl{$\bullet$ Lower bound:} The lower bound $\frac{n(n-1)}2$ for $n>2$ is attained by a series of \sa{} $\{\mathrsfs{V}_{n}\}$ found in~\cite[Theorem 4]{Gonze:2019}. The DFA $\mathrsfs{V}_n$ has $\{0,1,\dots,n-1\}$ as the state set and $\Sigma:=\{a_1,\dots,a_{n}\}$ as the input alphabet. The letter $a_n$ fixes all states except state 1, which it maps to 0, and for $1\le i \le n-1$, the letter $a_i$ fixes all states except states $i-1$ and $i$, which it swaps. Fig.~\ref{fig:rystsovmodified-n} shows a generic automaton from the series $\{\mathrsfs{V}_{n}\}$. That the DFAs $\mathrsfs{V}_n$ for $n>2$ belong to class \textbf{D5} is proved in~\cite[Lemma 6]{Ruszil:2023}.
\begin{figure}[hbt]
\begin{center}
\begin{tikzpicture}[->,>=latex,shorten >=1pt,auto,node distance=22mm,semithick]
  \tikzstyle{every state}=[minimum size=8mm,inner sep=0pt]

  \node[state] (A) {0};
  \node[state] (B) [right of=A] {1};
  \node[state] (C) [right of=B] {2};
  \node[state] (D) [right of=C] {$3$};
  \node (dots) [right of=D,xshift=-10mm] {$\cdots$};
  \node[state] (E) [right of=dots,xshift=-10mm] {$n{-}2$};
  \node[state] (F) [right of=E] {$n{-}1$};

  \path (B) edge[bend left=20] node[below] {$a_1,a_n$} (A)
        (A) edge[bend left=20] node[above] {$a_1$} (B)

        (C) edge[bend left=20] node[below] {$a_2$} (B)
        (B) edge[bend left=20] node[above] {$a_2$} (C)

        (D) edge[bend left=20] node[below] {$a_3$} (C)
        (C) edge[bend left=20] node[above] {$a_3$} (D)

        (F) edge[bend left=20] node[below] {$a_{n-1}$} (E)
        (E) edge[bend left=20] node[above] {$a_{n-1}$} (F);

  \path (A) edge[loop above,distance = 1cm] node[align=center] {\footnotesize $\Sigma{\setminus}\{a_1\}$} (A)
        (B) edge[loop above,distance = 1cm] node[align=center] {\footnotesize $\Sigma{\setminus}\{a_1,a_2\}$} (B)
        (C) edge[loop above,distance = 1cm] node[align=center] {\footnotesize $\Sigma{\setminus}\{a_2,a_3\}$} (C)
        (D) edge[loop above,distance = 1cm] node[align=center] {\footnotesize $\Sigma{\setminus}\{a_3,a_4\}$} (D)
        (E) edge[loop above,distance = 1cm] node[align=center] {\footnotesize $\Sigma{\setminus}\{a_{n{-}2},a_{n{-}1}\}$} (E)
        (F) edge[loop above,distance = 1cm] node[align=center] {\footnotesize $\Sigma{\setminus}\{a_{n{-}1}\}$} (F);
\end{tikzpicture}

\caption{The automaton $\mathrsfs{V}_n$}\label{fig:rystsovmodified-n}
\end{center}
\end{figure}

\textsl{$\bullet$ Version stamp:} Item added in August 2025.

\medskip

\noindent\textbf{D6. Automata with transitive permutation letters}
\nopagebreak

\textsl{$\bullet$ Definition:} For a DFA $\langle Q,\Sigma\rangle$ and $i=1,2\dots,|Q|$, let $\Sigma_i:=\{a\in\Sigma\mid |Q\setminus Q\dt a|=i\}$.  Class \textbf{D6} consists of DFAs $\langle Q,\Sigma\rangle$ such that $\Sigma=\Sigma_0\cup\Sigma_1$ and the DFA $\langle Q,\Sigma_0$ is \scn; that is, the permutation group on $Q$ generated by the permutations $p\mapsto p\dt a$, where $a\in\Sigma_0$, is transitive.

\textsl{$\bullet$ Upper bound:} $2n^2-7n+7$, see \cite[Theorem 4]{Zhu:2024}.

\textsl{$\bullet$ Method:} Enhanced extension method.

\textsl{$\bullet$ Lower bound:} $(n-1)^2$,  since the automata $\mC_n$ from the \v{C}ern\'y series lie in \textbf{D6}.

\textsl{$\bullet$ Version stamp:} Item added in September 2025.

\section*{Appendix}

\noindent\textbf{A2. One-cluster DFAs with a cycle of prime length: Lower bound}

Consider the DFA $\mD_{n,k}:=\langle\{0,1,\dots,n-1\},\{a,b\}\rangle$, with $k<n$ and the following action of the input letters:
\[
m\dt a:=\begin{cases}
0&\text{if }\ m=k-1,\\
n-k&\text{if }\ m=n-1,\\
m+1&\text{if }\ m\ne k-1,n-1;
\end{cases}
\qquad m\dt b:=m+1\!\!\pmod{n}.
\]
We prove that if $n$ and $k$ are coprime, then $\mD_{n,k}$ is a \san{} with \rl{} $k(n-2)+2$. The argument closely follows the proof of \cite[Theorem~6]{Ananichev&Volkov&Gusev:2013}, where the special case $k=n-1$ was considered (the DFA $\mathrsfs{D}_{n,n-1}$ appears as $\mD'_n$ in \cite{Ananichev&Volkov&Gusev:2013}).

It is easy to verify that if $n$ and $k$ are coprime, then the word $(ab^{k-1})^{n-2}ba$ of length $k(n-2)+2$ resets the automaton $\mD_{n,k}$.

The proof that $\mD_{n,k}$ has no shorter \sws{} makes use of the following elementary result:
\begin{lemma}[{\!\!\cite[Theorem~2.1.1]{RamirezAlfonsin:2005}}]
\label{lem:sylvester}
If $n,k$ are coprime positive integers, then $nk-n-k$ is the largest integer that is not expressible as a non-negative integer combination of $n$ and $k$.
\end{lemma}

Let $w$ be a \ssw\ for the automaton $\mD_{n,k}$. State $n{-}k$ is the only state in $\mD_{n,k}$ that serves as the common end of two distinct edges with the same label. Therefore, the minimality of $w$ implies that $w$ resets $\mD_{n,k}$ to $n{-}k$. Consequently, for every word $u\in\{a,b\}^*$, the word $uw$ also resets $\mD_{n,k}$ to $n{-}k$; in particular, $(n{-}k)\dt uw=n{-}k$. This equality implies that in the underlying graph of $\mD_{n,k}$,
there exists a directed cycle of any length $\ell\ge|w|$. The underlying graph of $\mD_{n,k}$ has exactly three simple directed cycles: one of length $n$ and two of length $k$, the latter being
\[
0\to 1\to\cdots\to k{-}1\to 0\ \text{ and }\ n{-}k\to n{-}k{+}1\to\cdots\to n{-}1\to n{-}k.
\]
Since every directed cycle is a concatenation of simple directed cycles, it follows that each $\ell\ge |w|$ must be a non-negative integer combination of $n$ and $k$. Lemma~\ref{lem:sylvester} then yields $|w|>nk-n-k$.

Now set $d:=|w|-(nk-n-k)$. Assume $d\le n-k+1$ and define
\[
q:=\begin{cases}n-k-d&\text{if $d\le n-k$},\\
k-1&\text{if $d=n-k+1$}.\end{cases}
\]
Viewing $q$ as a state of the automaton $\mD_{n,k}$, one has $q\dt w=n{-}k$ as the word $w$ sends every state to state $n{-}k$. Hence, in the underlying graph of $\mD_{n,k}$, there is a path $\pi$ from $q$ to $n{-}k$ labeled by $w$. Traversing the $d$ first edges of $\pi$ leads to state $n{-}k$. This is immediate if $q=n-k-d$, because in that case the $i$th of these $d$ edges is either $i\xrightarrow{a}i{+}1$ or $i\xrightarrow{b}i{+}1$, where $n-k-d\le i<n-k$. If $q=n-1$, note that the word $w$ starts with the letter~$a$ because $w$ is a \ssw\ for the automaton $\mD_{n,k}$ and the letter $b$ acts as a permutation. Hence the first edge of the path~$\pi$ is $k{-}1\xrightarrow{a}0$, and each of the consequent $d-1$ edges is either $i\xrightarrow{a}i{+}1$ or $i\xrightarrow{b}i{+}1$, where $0\le i<n-k$. Therefore, after the first $d$ edges, the path $\pi$ enters a cycle of length $|w|-d=nk-n-k$, but Lemma~\ref{lem:sylvester} implies that no cycle of this length may exist in the underlying graph of $\mD_{n,k}$, a contradiction.

We have thus proved that $d\ge n-k+2$, and hence
\[
|w|=nk-n-k+d\ge nk-n-k+n-k+2=k(n-2)+2,
\]
as claimed.

To get the lower bound $(n-2)(n-n^{0.525})+2$ for the \rl{} of \sa{} in class \textbf{A2} with sufficiently large composite number~$n$ of states, let $p$ be the greatest prime less than $n$. It is known \cite[Theorem~1]{Bakeretal:2001} that $p>n-n^{0.525}$ for sufficiently large $n$. Clearly, $n$ and $p$ are coprime, and the fact just established applies to the DFA $\mD_{n,p}$. Hence, $\rt(\mathrsfs{D}_{n,p})=p(n-2)+2\ge (n-2)(n-n^{0.525})+2$ for all sufficiently large $n$.

To get a lower bound  the \rl{} of \sa{} in class \textbf{A2} with any composite number of states, one can employ Bertand's postulate, a classic fact of number theory, which ensures that the greatest prime $p$ less than $n$ satisfies $p>\frac{n}2$. This yields $\rt(\mathrsfs{D}_{n,p})=p(n-2)+2\ge\frac{n^2}2$.

\medskip

\noindent\textbf{A10. Binary DFAs with a simple idempotent letter: Upper bound}

Let $\mA=\langle Q,\{a,b\}\rangle$ be a \san{} such that $|Q|=n$ and the letter $a$ is a simple idempotent. We prove that $\mA$ has a \sw{} of length at most $(n-1)^2$.

Let $e$ stand for the only state in $Q\setminus Q\dt a$. We say that a state $q\in Q$ is $b$-cyclic if $q=q\dt b^\ell$ for some positive integer $\ell$. A $b$-cycle of $\mA$ is any set of states of the form $\{q\dt b^k \mid q \text{ is a $b$-cyclic state, $k$ is a nonnegative integer}\}$. The letter $b$ acts as a cyclic permutation on each $b$-cycle.

If a $b$-cycle $C$ excludes the state $e$, then $C\subseteq Q\dt a$ whence $C\dt a=C$ as $a$ fixes every state in $Q\dt a$. Besides, $C\dt b=C$ as $b$ acts as a cyclic permutation on $C$. Hence, $C\dt w=C$ for every word $w\in\{a,b\}^*$. On the other hand, letting $w$ be a \sw\ for $\mA$, we get $|C\dt w|=1$. We see that $C$ must consist of just one state $z$ such that $z\dt a=z\dt b=z$ so $z$ is a zero of $\mA$. Every $n$-state \san\ with a zero state has a \sw\ of length at most $\frac{n(n-1)}2\le(n-1)^2$; see item B1.

We may therefore assume that every $b$-cycle contains the state $e$. This implies that $\mA$ has a unique $b$-cycle as distinct $b$-cycles are pairwise disjoint. Let $C$ be this $b$-cycle and $m:=|C|$.

For any $q\in Q\setminus C$, consider the sequence $q,q\dt b,\dots,q\dt b^{n-m}$. If all $n-m+1$ states in this sequence are distinct, then one of them lies in $C$, and if $q\dt b^k=q\dt b^\ell$ for some $0\le k<\ell\le n-m$, then $q\dt b^k$ is a $b$-cyclic state whence it lies in $C$ as $C$ is the only $b$-cycle in $\mA$. Clearly, if $q\dt b^k\in C$ for some $k$, then $q\dt b^\ell\in C$ for $\ell\ge k$. Hence $q\dt b^{n-m}\in C$ for every state $q\in Q$. If $m=1$, then the word $b^{n-1}$ synchronizes $\mA$ so assume $m>1$ for the rest of the proof.

Let $d:=e\dt a$. Consider two cases.

\textbf{Case 1:} $d\in C$. Then $C\dt a\subset C$. We may therefore consider the subautomaton $\mC=(C,\{a,b\})$. Any \sw\ for $\mA$ synchronizes $\mC$, as well, so $\mC$ is a \san. The letter $b$ acts as a cyclic permutation on the state set of $\mC$, and by the result from item A1, $\mC$ has a \sw\ $u$ of length at most $(m-1)^2$. Since $Q\dt b^{n-m}\subseteq C$, the word $b^{n-m}u$ of length at most $n-m+(m-1)^2\le(n-1)^2$ synchronizes $\mA$.

\textbf{Case 2:} $d\notin C$. Let $k>0$ be the least integer such that $r:=d\dt b^k\in C$ and let $D:=\{d,d\dt b,\dots,d\dt b^{k-1}\}$. See Fig.~\ref{fig:case 2} for illustration.
\begin{figure}[ht]
\begin{center}
\makebox[0pt][c]{\hspace{2.5cm}%
 \begin{tikzpicture}[>=latex, shorten >=1pt, auto]
  \def \m {6} 
  \def \radius {2.5cm}
  \foreach \i in {1,...,\m} {
    \node[circle, draw, minimum size=8mm] (C\i) at ({360/\m * (\i-1)}:\radius) {};
  }

\draw[-latex] (C1) -- (C2) node[midway,sloped,below] {$b$};
\draw[-latex] (C2) -- (C3) node[midway,sloped,below] {$b$};
\draw[-latex] (C3) -- (C4) node[midway,sloped,below] {$b$};
\draw[-latex] (C4) -- (C5) node[midway,sloped,above] {$b$};
\draw[-latex,dashed] (C5) -- (C6) node[midway,sloped,above] {$b^{m-5}$};
\draw[-latex] (C6) -- (C1) node[midway,sloped,above] {$b$};

  \node at (C1) {$e$};
  \node at (C3) {$r$};

  \node[circle, draw, minimum size=8mm, left=4cm of C3] (d0) {$d$};

  \node[circle, draw, minimum size=8mm, midway, left=6cm of d0] (d1) {};
  \node[circle, draw, minimum size=8mm, midway, left=3.5cm of d1] (d2) {};

  \draw[-latex] (d0) -- (d1) node[midway,right] {$b$};
  \draw[-latex] (d1) -- (d2) node[midway,above] {$b$};
  \draw[-latex, dashed] (d2) -- (C3) node[midway,left] {$b^{k-2}$};

  \draw[-latex] (C1) .. controls +(3,4.5) and +(0,2) .. (d0) node[midway,above] {$a$};

\end{tikzpicture}}
\end{center}
\caption{The subautomaton $\langle C\cup D,\{a,b\}\rangle$ of $\mA$. Loops labeled $a$ at all states except $e$ are omitted for readability}\label{fig:case 2}
\end{figure}

 For any state $q\in C\cup D$, define its distance to $r$ as the least nonnegative integer $\dist(q,r)$ such that $q\dt b^{\dist(q,r)}=r$. In particular, $\dist(d,r)=k$. Clearly,
\begin{equation}\label{eq:distance+b}
\dist(q\dt b,r)=\begin{cases}
\dist(q,r)-1&\text{if $q\ne r$},\\
m-1 &\text{if $q=r$}.
\end{cases}
\end{equation}
Let $\ell:=\dist(e,r)$. Since $a$ fixes all the states in $C\cup D$ except for $e$, we have
\begin{equation}\label{eq:distance+a}
\dist(q\dt a,r)=\begin{cases}
\dist(q,r)&\text{if $q\ne e$},\\
\dist(q,r)+k-\ell&\text{if $q=e$}.
\end{cases}
\end{equation}
Let $g$ stand for the greatest common divisor of $m$ and $k-\ell$. A straightforward induction which steps are based on \eqref{eq:distance+b} and \eqref{eq:distance+a} shows that
\[
\dist(q,r)-\dist(q',r)\equiv \dist(q\dt w,r)-\dist(q'\dt w,r)\!\!\pmod g
\]
for all $q,q'\in C\cup D$ and every word $w\in\{a,b\}^*$. Choosing the states $e$ and $e\dt b$ for $q$ and $q'$, respectively, and a \sw\ of the automaton $\mA$ for $w$, we get
\[
\dist(e,r)-\dist(e\dt b,r)\equiv \dist(e\dt w,r)-\dist((e\dt b)\dt w,r)=0\!\!\pmod g,
\]
since $e\dt w=(e\dt b)\dt w$. On the other hand, \eqref{eq:distance+b} implies that
\[
\dist(e,r)-\dist(e\dt b,r)\equiv 1\!\!\pmod m
\]
whence $1\equiv 0\!\!\pmod g$. This only possible if $g=1$. We have thus proved that the numbers $k-\ell$ and $m$ are coprime.

Now let $v:=\begin{cases} a &\text{if $\ell=0$},\\ b^{m-\ell}a &\text{if $\ell>0$} \end{cases}$. We then have $r\dt vb^k=d\dt b^k=r$. For any $q\in C\setminus\{r\}$, there exists a unique $t$ with $0<t<m-1$ such that $\dist(q,r)\equiv t(k-\ell)\!\!\pmod m$---this follows from the just established fact that $k-\ell$ and $m$ are coprime. We have $q\dt b^{m-\ell}\ne e$ whence $q\dt v=q\dt b^{m-\ell}$ and $q\dt vb^k=q\dt b^{m+k-\ell}$. Using \eqref{eq:distance+b}, we compute
\[
\dist(q\dt vb^k,r)\equiv (t-1)(k-\ell)\!\!\pmod m.
\]
In particular, $q\dt vb^k=r$ if $t=1$. Hence $q\dt (vb^k)^{m-1}=r$ for all $q\in C$.  Since $Q\dt b^{n-m}\subseteq C$, the word $b^{n-m}(vb^k)^{m-1}$ resets $\mA$. The length of this word is
\[
\begin{cases}
(k+1)(m-1)+n-m &\text{if $\ell=0$},\\
(m-\ell+k+1)(m-1)+n-m &\text{otherwise.}
\end{cases}
\]
Since $m+k\le n$, we have both $k+1\le n$ and $m-\ell+k+1\le n$ so the length of $b^{n-m}(vb^k)^{m-1}$ does not exceed $n(m-1)+n-m=(n-1)m\le(n-1)^2$.\qed

\end{document}